\documentclass[
aps,
prd,
reprint,
superscriptaddress,
nofootinbib,
longbibliography,  floatfix
]{revtex4-2}

\usepackage{booktabs}
\usepackage{graphicx}
\usepackage{xcolor}
\usepackage{microtype}

\usepackage{amsmath}
\usepackage{amssymb}
\usepackage{mathrsfs}
\usepackage{bm}
\usepackage{array}[=2016-10-06]
\usepackage{tabularx}
\usepackage{float}

\usepackage[colorlinks=true,
citecolor=blue!55!black,
linkcolor=blue!55!black,
urlcolor=blue!55!black]{hyperref}

\graphicspath{{figures/}}

\newcommand{\dd}{\mathop{}\!\mathrm{d}}

\begin{document}
\title{Novel topological subclasses of a magnetic black hole in causal nonlinear electrodynamics}

\author{Qi-Qi Liang}
\affiliation{College of Physics, Guizhou University, Guiyang 550025, People’s Republic of China}

\author{De-jiang Yin}
\affiliation{College of Physics, Guizhou University, Guiyang 550025, People’s Republic of China}

\author{Yu-Die Wan}
\affiliation{College of Physics, Guizhou University, Guiyang 550025, People’s Republic of China}

\author{Zheng-Wen Long}
\email[Corresponding author: ]{zwlong@gzu.edu.cn}
\affiliation{College of Physics, Guizhou University, Guiyang 550025, People’s Republic of China}


\begin{abstract}
By treating black-hole states as topological defects in thermodynamic parameter space,  we investigate a static, asymptotically flat magnetic black holes in Einstein gravity coupled to causal nonlinear electrodynamics. Using the generalized off-shell free energy and Duan's $\phi$-mapping topological current theory, we analyze the physical single-horizon sector in the canonical ensemble. We identify four representative defect-curve profiles, including realizations of the conjectured $W^{0+\leftrightarrow1-}$ class and $\overline{W}^{1-}$ subclass, as well as a new thermodynamic topological subclass, $\overline{W}^{0+\leftrightarrow1-}$. This new subclass features paired generation and annihilation points, with unstable innermost and outermost branches. As the auxiliary inverse temperature increases, the system evolves from two equilibrium states to three and then back to one, with an unstable small-black-hole defect entering through the physical lower boundary and changing the global topological number from $W=0$ to $W=-1$. This boundary-induced change signals a thermodynamic topological phase transition. These results show that the same inverse-temperature endpoint behavior and winding-number sequence from small to large black holes can accompany different defect evolution, highlighting the role of the physical horizon boundary in thermodynamic
topological classification. More broadly, this causal
nonlinear-electrodynamics black-hole family provides a setting for
discovering further defect structures and thermodynamic topological
phase transitions.

\end{abstract}

\maketitle
\section{Introduction}

Black hole thermodynamics describes the thermal properties of gravitational
solutions through quantities such as the Hawking temperature, entropy,
free energy, and thermodynamic response functions
\cite{1973PhRvD...7.2333B,1975CMaPh..43..199H,
1977PhRvD..15.2752G,1983CMaPh..87..577H,
1986PhRvD..33.2092Y,1990PhRvD..42.3376B,
1993PhRvD..48.3427W}.
The Euclidean approach and the analysis of thermodynamic ensembles
provide the foundations for studying black hole equilibrium and phase
behavior, including the role of boundary conditions and ensemble
dependence
\cite{1983CMaPh..87..577H,1986PhRvD..33.2092Y,
1990PhRvD..42.3376B}.
Thermodynamic topology characterizes equilibrium black hole states
as zero points of a thermodynamic vector field constructed from
the generalized off-shell free energy. The $\phi$-mapping current formalism supplies the winding-number
description of isolated zeros
\cite{DuanGe1979,1998NuPhB.514..705D}.
Wei, Liu, and Mann applied this formalism to black-hole
thermodynamic defects defined by zero points of an off-shell
vector field \cite{2022PhRvL.129s1101W}.
Each isolated thermodynamic defect carries a local winding number, whose
sign distinguishes locally stable and unstable thermodynamic branches.
The sum of the winding numbers at a given auxiliary inverse temperature
defines the global topological number $W$. The topological analysis of black-hole critical points was developed
in Ref.~\cite{2022PhRvD.105j4003W}.\textit{ Wei }et al. subsequently
formulated four universal classes of black-hole thermodynamic defects:
$W^{1-}$, $W^{0+}$, $W^{0-}$, and $W^{1+}$
\cite{2024PhRvD.110h1501W}.
Subsequent studies extended this classification by identifying additional
classes, subclasses, and candidate structures, including
$W^{0-\leftrightarrow1+}$, $\overline W^{1+}$, and related
higher-order branch structures
\cite{2025PhRvD.111f1501W}.

Rotating black holes have supplied further thermodynamic topological
structures
\cite{2025PhLB..86039163Z,2025PhRvD.112l4024A}.
Static accelerating black holes
\cite{2023PhRvD.108h4041W} and static multi-charge AdS
black holes in gauged supergravity
\cite{2024JHEP...06..213W} have also been studied.
Higher-curvature and modified-gravity examples include Lovelock,
Gauss--Bonnet, $f(R)$, and Ho\v{r}ava--Lifshitz black holes
\cite{2023PhRvD.107f4015B,2023PhRvD.107f4023L,
2025PhRvD.111f4006J,2024PTEP.2024d3E01H,
2024NuPhB100616649H,2025EPJC...85.1386C}.
Related investigations of BTZ and quantum BTZ black holes
have revealed further thermodynamic topological structures
\cite{2024Symm...16.1577D,2024PhRvD.110b4054W}.
Related topological studies have also addressed thermodynamic
critical points in hairy Lovelock and dyonic
Einstein--Gauss--Bonnet black holes
\cite{2024EPJC...84.1251Z,2024EPJP..139..806S}.

In particular, \textit{Wu} et al. identified the $W^{0+\leftrightarrow1-}$ class, characterized by stable innermost and unstable outermost branches, with one more generation point (GP) than annihilation point (AP), and the $\overline W^{1-}$ subclass, featuring unstable innermost and outermost branches with equal numbers of GPs and APs \cite{2025PhRvD.111f1501W}. Both exhibit a finite positive inverse-temperature asymptotics behavior at the minimum admissible horizon radius and $\beta(\infty)=\infty$. The $\overline W^{1-}$ subclass has also been realized in slowly rotating magnetic black holes in nonlinear electrodynamics and magnetic GMGHS black holes \cite{1988NuPhB.298..741G,1991PhRvD..43.3140G,2026arXiv260916568Z,2026arXiv260920898Y}.

These developments show that the global topological number alone does not
fully specify the structure of thermodynamic defects. The endpoint
behavior, radial ordering of winding numbers, and evolution of defect
branches provide additional information for distinguishing different
thermodynamic topological realizations
\cite{2024PhRvD.110h1501W,2025PhRvD.111f1501W}.
Nevertheless, different orderings between the boundary inverse
temperature and the extrema of the defect curve can generate distinct
defect evolutions. A variation of the global topological number
may occur when a defect enters or leaves through the boundary of the
physical thermodynamic parameter space, whereas interior generation and
annihilation processes involve opposite-winding pairs and preserve the
global topological number
\cite{2026arXiv260707364W}.

Born--Infeld electrodynamics is an early nonlinear extension of
Maxwell theory \cite{1933Natur.132.1004B}.
Subsequent nonlinear electrodynamics (NLED) models have led to
various regular and AdS black hole solutions
\cite{1998PhRvL..80.5056A,1999PhLB..464...25A,
2004PhLB..595..484K}.
Recently, Fathi et al. constructed a static
magnetic black hole solution in Einstein--NLED gravity with
a Maxwell weak-field limit, finite magnetic self-energy,
causal electromagnetic propagation, and sufficient exterior
stability conditions \cite{2026arXiv260806639F}.
In the sector $M_0\geq0$, the metric function is monotonic
and admits a single event horizon, with the physical lower
horizon radius $r_m$ determined by $M_0(r_m)=0$.
This horizon domain is used in the present thermodynamic
topology analysis.

In this work, we investigate the thermodynamic topology of this magnetic black hole in the canonical ensemble at fixed physical magnetic charge. We derive the thermodynamic quantities along the admissible horizon family and construct the thermodynamic vector field from the generalized off-shell free energy. The zero point condition determines the defect curve, which is identified with the inverse Hawking temperature through the first law. The inverse temperature approaches a finite positive value at the lower horizon boundary $r_m$ and diverges as the horizon radius tends to infinity. We identify four representative thermodynamic branch structures, labeled Types I--IV. Type I realizes the previously proposed $W^{0+\leftrightarrow1-}$ pattern, whereas Types II and III provide two realizations of the $\overline W^{1-}$ subclass. 
Type IV shares the radial winding sequence $[-,+,-]$ and one
generation point--annihilation point pair with Type III, but has the reversed ordering
$\hat\beta_{\min}^{\mathrm{out}}<\hat\beta_m<\hat\beta_{\max}$.
As $\hat\tau$ increases, its nonempty defect configurations
evolve as $[+,-]\to[-,+,-]\to[-]$, with the entry of an
unstable defect through the lower horizon boundary changing
$W:0\to-1$.
This branch structure motivates the notation $\overline W^{0+\leftrightarrow1-}$ as a thermodynamic topological subclass candidate, and the present Einstein--NLED family provides an explicit realization within the explored parameter region.

The remainder of this paper is organized as follows. In Sec.~\ref{sec:2}, we review the magnetic black hole solution and its admissible horizon domain. Section~\ref{sec:3} presents the thermodynamic quantities, generalized off-shell free energy, and thermodynamic vector field. In Sec.~\ref{sec:4}, we examine the inverse-temperature endpoints and introduce dimensionless variables. Section~\ref{sec:5} analyzes the parameter dependence, thermodynamic branches, winding numbers, and boundary contributions. Section~\ref{sec:6} contains the discussion and conclusions.

\section{Magnetic black hole solution and physical horizon domain}
\label{sec:2}
We consider the static magnetic black hole solution of Fathi, Guzm\'an,
and Villanueva in Einstein gravity coupled to nonlinear electrodynamics
(NLED)~\cite{2026arXiv260806639F}. In units $G=c=1$, the action is
\begin{equation}
I=\frac{1}{16\pi}\int d^4x\sqrt{-g}\,
\left[R-4\mathcal L(\mathcal F)\right],
\end{equation}
where $\mathcal F=F_{\mu\nu}F^{\mu\nu}/4$ and
\begin{align}
\mathcal L(\mathcal F)
={}&\frac{3(1-\xi)}{2\beta}
\left[(1+\beta\mathcal F)^{2/3}-1\right]
\nonumber\\
&+\frac{2\xi}{\sigma\beta}
\left[(1+\sigma\beta\mathcal F)^{1/2}-1\right].
\end{align}
Here $\beta>0$ and $\sigma>0$ are NLED parameters. We consider the
mixed sector $0<\xi<1$ with nonzero magnetic charge $Q_m$.
The Lagrangian has the Maxwell weak-field limit and finite magnetic
self-energy. 
NLED wave propagation and causality have been analyzed in
Refs.~\cite{1970JMP....11..941B,2000PhRvD..61d5001N,
2001PhRvD..63f4006G}. Sufficient perturbative stability
conditions for magnetic black holes are given in
Ref.~\cite{2020PhRvD.101l4026N}.
The present solution satisfies the relevant conditions
\cite{2026arXiv260806639F}.

For the magnetic field $F_{\theta\varphi}=Q_m\sin\theta$, the
invariant is $\mathcal F=Q_m^2/(2r^4)$. The metric and mass function
satisfy
\begin{align}
ds^2&=-f(r)dt^2+\frac{dr^2}{f(r)}+r^2d\Omega_2^2,
\nonumber\\
f(r)&=1-\frac{2m(r)}{r},\qquad
m'(r)=r^2\mathcal L\!\left(\frac{Q_m^2}{2r^4}\right).
\end{align}
Writing the electromagnetic contribution with $m_{\rm em}(0)=0$ gives
\begin{align}
m(r)&=M_0+m_{\rm em}(r),\nonumber\\
m_{\rm em}(r)&=\int_0^r s^2\mathcal L\!\left(
\frac{Q_m^2}{2s^4}\right)ds,
\nonumber\\
M&=M_0+E_{\rm em},\qquad
E_{\rm em}=m_{\rm em}(\infty)>0.
\end{align}
Here $M_0$ is the Schwarzschild mass parameter and $M$ is the ADM mass.
At $M_0=0$, the ADM mass remains $M=E_{\rm em}>0$.

We work in the single-horizon sector $M_0\geq0$ of the original
solution. Introducing $D(r)=m(r)-rm'(r)$, one obtains
\begin{equation}
f'(r)=\frac{2D(r)}{r^2},\qquad
D'(r)=2r^2\left(2\mathcal F\mathcal L_{\mathcal F}
-\mathcal L\right)>0.
\label{eq:D}
\end{equation}
Since $D(0^+)=M_0\geq0$, the function $D(r)$ is positive for $r>0$,
and hence $f'(r)>0$. For $0<\xi<1$, the metric function approaches
$-\infty$ as $r\to0^+$ and unity as $r\to\infty$. The solution
therefore has a unique positive event horizon in this sector
\cite{2026arXiv260806639F}.

The horizon condition $f(r_+)=0$ determines the mass parameter as
\begin{equation}
M_0(r_+)=\frac{r_+}{2}-m_{\rm em}(r_+),\qquad
M_0(r_m)=0.
\label{eq:M0}
\end{equation}
Because $dM_0/dr_+=r_+f'(r_+)/2>0$, the horizon radius increases
monotonically with $M_0$. Thus the selected family has the horizon domain
\begin{equation}
M_0\in[0,\infty)\quad\Longleftrightarrow\quad
r_+\in[r_m,\infty).
\label{eq:domain}
\end{equation}
The radius $r_m$ is the lower horizon endpoint corresponding to the
$M_0=0$ black hole. Thermodynamic defects are counted in the interior
$r_+>r_m$. Throughout the thermodynamic analysis, $Q_m$, $\beta$,
$\sigma$, and $\xi$ are held fixed, while $M_0$ and $r_+$ vary.

\section{Thermodynamics and the thermodynamic vector field}
\label{sec:3}
For fixed $(Q_m,\beta,\sigma,\xi)$, the horizon condition
$f(r_+)=0$ gives $m(r_+)=r_+/2$. The ADM mass along the
one-horizon family is therefore
\begin{equation}
M(r_+)=\frac{r_+}{2}
+\int_{r_+}^{\infty}s^2\mathcal L\!\left(
\frac{Q_m^2}{2s^4}\right)\dd s .
\label{eq:mass-horizon}
\end{equation}
The entropy obeys the Einstein area law,
\begin{equation}
S(r_+)=\pi r_+^2.
\label{eq:entropy}
\end{equation}
The Hawking temperature, obtained from the surface gravity at the
event horizon, is
\begin{equation}
T_H(r_+)=\frac{f'(r_+)}{4\pi}
=\frac{1-2r_+^2\mathcal L(\mathcal F_+)}{4\pi r_+}>0.
\label{eq:temperature}
\end{equation}
Here $\mathcal F_+=Q_m^2/(2r_+^4)$. Differentiating
Eq.~\eqref{eq:mass-horizon} at fixed magnetic charge
and fixed NLED parameters yields
\begin{equation}
M'(r_+)=\frac12-r_+^2\mathcal L(\mathcal F_+)
=T_H(r_+)S'(r_+),
\label{eq:firstlaw}
\end{equation}
which is the first law along the thermodynamic family,
$\dd M=T_H\dd S$.

We introduce an independent auxiliary inverse temperature $\tau>0$
and define the generalized off-shell free energy
\begin{equation}
F_{\rm off}(r_+,\tau)=M(r_+)-\frac{S(r_+)}{\tau}.
\end{equation}

We use the $\phi$-mapping current framework
\cite{1998NuPhB.514..705D} and the off-shell black-hole
thermodynamic construction
\cite{2022PhRvL.129s1101W,2023JHEP...01..102F}.
The resulting two-component vector field is
\begin{align}
\boldsymbol\phi&=(\phi^{r_+},\phi^\Theta),\nonumber\\
\phi^{r_+}&=\left.\frac{\partial F_{\rm off}}{\partial r_+}
\right|_{\tau,Q_m,\beta,\sigma,\xi}\nonumber\\
&=\frac12-r_+^2\mathcal L(\mathcal F_+)
-\frac{2\pi r_+}{\tau},\nonumber\\
\phi^\Theta&=-\frac{\cos\Theta}{\sin^2\Theta},
\qquad 0<\Theta<\pi.
\label{eq:off}
\end{align}
At fixed $\tau$, the field is defined on the thermodynamic parameter
space $(r_+,\Theta)$, with $r_+>r_m$. Its zero points satisfy
$\Theta=\pi/2$ and $\phi^{r_+}=0$. The radial zero point condition
gives the defect curve
\begin{equation}
\tau=\tau_0(r_+)=\frac{S'(r_+)}{M'(r_+)}
=\frac{4\pi r_+}{1-2r_+^2\mathcal L(\mathcal F_+)}.
\end{equation}
Using Eq.~\eqref{eq:firstlaw}, this curve is identified with the
inverse Hawking temperature,
$\tau_0(r_+)=\beta_H(r_+)\equiv 1/T_H(r_+)$.
Explicitly,
\begin{align}
\beta_H(r_+)=4\pi r_+\Bigg\{1
&-\frac{3(1-\xi)r_+^2}{\beta}
\left[\left(1+\frac{\beta Q_m^2}{2r_+^4}\right)^{2/3}-1\right]
\nonumber\\
&-\frac{4\xi r_+^2}{\sigma\beta}
\left[\left(1+\frac{\sigma\beta Q_m^2}{2r_+^4}\right)^{1/2}-1\right]
\Bigg\}^{-1}.
\label{eq:tauexplicit}
\end{align}
Away from the zero point condition, $\tau$ remains an independent
parameter of $F_{\rm off}$.

\subsection{Local winding numbers and turning points}

For a fixed regular value of $\tau$, let
$Z_i=(r_i,\pi/2)$ denote an isolated simple zero of the
thermodynamic vector field. We use the orientation $(r_+,\Theta)$.
Since $\partial_\Theta\phi^\Theta|_{\Theta=\pi/2}=1$, the Jacobian
at $Z_i$ is
\begin{align}
J_i&=\left.\det\frac{\partial(\phi^{r_+},\phi^\Theta)}
{\partial(r_+,\Theta)}\right|_{Z_i}\nonumber\\
&=\left.\frac{\partial^2F_{\rm off}}
{\partial r_+^2}\right|_{Z_i}
=2\pi r_iT_H'(r_i).
\label{eq:jac}
\end{align}
For a simple zero the Hopf index is unity, and the winding number is
\begin{equation}
w_i=\operatorname{sgn}J_i
=\operatorname{sgn}T_H'(r_i)
=-\operatorname{sgn}\beta_H'(r_i).
\end{equation}
Thus $w_i=+1$ ($w_i=-1$) corresponds to a local minimum
(maximum) of the off-shell free energy in the radial direction,
identifying a locally stable (unstable) thermodynamic branch.
The global topological number on a regular inverse-temperature slice is
\begin{equation}
W(\tau)=\sum_{\,r_i\in(r_m,\infty)}w_i.
\end{equation}

The turning points of the defect curve are determined by
\begin{align}
\mathcal P(r_+)
&=-1+8r_+^2\mathcal F_+
\mathcal L_{\mathcal F}(\mathcal F_+)
-2r_+^2\mathcal L(\mathcal F_+),\nonumber\\
T_H'(r_+)&=\frac{\mathcal P(r_+)}{4\pi r_+^2},
\qquad
\beta_H'(r_+)=-\frac{T_H'(r_+)}{T_H(r_+)^2}.
\label{eq:heat}
\end{align}
Consequently, $w_i=\operatorname{sgn}\mathcal P(r_i)$ for each
simple zero. An interior turning point satisfies
$\mathcal P(r_+)=0$ and $J=0$. For a simple turning point,
$\mathcal P'(r_+)\ne0$, a local minimum (maximum) of
$\beta_H$ corresponds to a generation (annihilation) point as
$\tau$ increases.

\section{Dimensionless defect curve and endpoint behavior}
\label{sec:4}
For the thermodynamic family at fixed $(Q_m,\beta,\sigma,\xi)$,
we introduce the length scale $b$ and the dimensionless variables
\begin{align}
b^4&=\frac{\beta Q_m^2}{2},\qquad
\hat r_+=\frac{r_+}{b},\qquad
\hat r_m=\frac{r_m}{b},\nonumber\\
\lambda&=\frac{|Q_m|}{\sqrt{\beta}},\qquad
\hat\tau=\frac{\tau}{4\pi b},\qquad
\hat\beta_H=\frac{\beta_H}{4\pi b}.
\label{eq:scales}
\end{align}
Here $b>0$ and $b^2/\beta=\lambda/\sqrt{2}$.
The parameter
$\beta$ is the NLED coupling, whereas $\beta_H$ denotes the inverse
Hawking temperature. With $y=\hat r_+^{-4}=\beta\mathcal F_+$,
we define the dimensionless NLED Lagrangian
\begin{align}
\ell(y)\equiv\beta\mathcal L(y/\beta)
={}&\frac{3(1-\xi)}{2}\left[(1+y)^{2/3}-1\right]
\nonumber\\
&+\frac{2\xi}{\sigma}\left[(1+\sigma y)^{1/2}-1\right].
\end{align}
The thermodynamic vector field and the defect curve can then be
written in terms of
\begin{align}
A(\hat r_+)
&=1-\sqrt{2}\lambda\hat r_+^2\ell(\hat r_+^{-4}),
\nonumber\\
\phi^{r_+}
&=\frac12\left[A(\hat r_+)-\frac{\hat r_+}{\hat\tau}\right],
\nonumber\\
\hat\beta_H(\hat r_+)
&=\frac{\hat r_+}{A(\hat r_+)},
\nonumber\\
\mathcal P(\hat r_+)
&=-1+\sqrt{2}\lambda\hat r_+^2
\left[4\hat r_+^{-4}\ell_y(\hat r_+^{-4})
-\ell(\hat r_+^{-4})\right],
\nonumber\\
\frac{\dd\hat\beta_H}{\dd\hat r_+}
&=-\frac{\mathcal P(\hat r_+)}{A(\hat r_+)^2},
\qquad \ell_y=\frac{\dd\ell}{\dd y}.
\label{eq:ell}
\end{align}
At a zero point of the thermodynamic vector field,
$\hat\tau=\hat\beta_H(\hat r_+)$. Since
$A(\hat r_+)=4\pi r_+T_H(r_+)>0$ on the selected
one-horizon family, the interior extrema of the defect curve
satisfy $\mathcal P(\hat r_+)=0$. The sign of $\mathcal P$
determines the winding number of each simple thermodynamic defect.

The lower endpoint follows from the horizon relation at $M_0=0$.
Defining
\begin{equation}
\mathcal J(x)=\int_0^x u^2\ell(u^{-4})\dd u,
\end{equation}
the Schwarzschild mass parameter becomes
\begin{align}
\frac{M_0}{b}
&=\frac{\hat r_+}{2}
-\frac{\lambda}{\sqrt{2}}\mathcal J(\hat r_+),
\nonumber\\
\lambda
&=\frac{\hat r_m}{\sqrt{2}\,\mathcal J(\hat r_m)}.
\label{eq:endpoint}
\end{align}
The second relation determines $\lambda$ from the endpoint
radius for fixed $(\xi,\sigma)$. Thus $(\xi,\sigma,\hat r_m)$
can be used to specify a dimensionless thermodynamic family;
$\lambda$ remains fixed as $\hat r_+$ varies over
$(\hat r_m,\infty)$.

The Maxwell weak-field limit gives $\ell(y)=y+O(y^2)$ as
$y\to0$. The inverse-temperature endpoints are therefore
\begin{align}
0<\hat\beta_m
&\equiv\lim_{\hat r_+\to\hat r_m^+}
\hat\beta_H(\hat r_+)
=\frac{\hat r_m}{A(\hat r_m)}<\infty,
\nonumber\\
\hat\beta_H(\hat r_+)
&=\hat r_+ +\frac{\sqrt{2}\lambda}{\hat r_+}
+O(\hat r_+^{-3}),\qquad \hat r_+\to\infty.
\label{eq:limits}
\end{align}
The turning-point function has the asymptotic form
\begin{equation}
\mathcal P(\hat r_+)
=-1+\frac{3\sqrt{2}\lambda}{\hat r_+^2}
+O(\hat r_+^{-6}),\qquad \hat r_+\to\infty.
\end{equation}
Hence $\hat\beta_H(\infty)=\infty$, and the large-radius
thermodynamic branch has winding number $w=-1$.
At the lower endpoint, the adjacent branch has
$w=\operatorname{sgn}\mathcal P(\hat r_m)$ when
$\mathcal P(\hat r_m)\ne0$.
Its winding number and the ordering of the interior extrema relative
to $\hat\beta_m$ determine the distinct branch structures examined in
Sec.~V.

\section{Parameter dependence and thermodynamic defect structures}
\label{sec:5}
The admissible thermodynamic branches are determined by the zeros of
$\mathcal P(\hat r_+)$ on $\hat r_+>\hat r_m$ and by the ordering of the
corresponding inverse temperatures relative to
$\hat\beta_m=\hat\beta_H(\hat r_m)$.  Since
$\mathrm{d}\hat\beta_H/\mathrm{d}\hat r_+=-\mathcal P/A^2$,
$\mathcal P>0$ and $\mathcal P<0$ identify branches carrying winding
numbers $w=+1$ and $w=-1$, respectively.  A zero of $\mathcal P$ in the
interior is a turning point of the defect curve
$\hat\tau=\hat\beta_H(\hat r_+)$.  We characterize the resulting
branch structures by their winding sequences, turning points, and
finite-radius endpoint ordering.

\subsection{Parameter regions and classification criteria}

For simple interior turning points, define
\begin{equation}
 N_{\mathrm{ext}}=\#\bigl\{\hat r_+>\hat r_m:
 \mathcal P(\hat r_+)=0,\ \mathcal P'(\hat r_+)\neq0\bigr\}.
 \label{eq:Next}
\end{equation}
A minimum of $\hat\beta_H$ is a generation point (GP) as
$\hat\tau$ increases, while a maximum is an annihilation point (AP),
following the defect evolution picture used in thermodynamic topology
classifications
\cite{2025PhRvD.111f1501W}.
The two-turning-point profiles have radii $\hat r_a<\hat r_b$, at
which we define
\begin{align}
 \hat\beta_{\max}&=\hat\beta_H(\hat r_a),\nonumber\\
 \hat\beta_{\min}^{\mathrm{out}}&=\hat\beta_H(\hat r_b),\nonumber\\
 \Delta\hat\beta&=\hat\beta_{\min}^{\mathrm{out}}-\hat\beta_m.
 \label{eq:delta}
\end{align}
Thus $\Delta\hat\beta$ distinguishes two profiles that have the
same radial winding sequence $[-,+,-]$ but different defect contents
on the interval between the boundary and the outer minimum.

The changes in branch structure are associated with
\begin{equation}
\mathcal P(\hat r_m)=0,\qquad
\mathcal P(\hat r_c)=\mathcal P'(\hat r_c)=0,
\quad \hat r_c>\hat r_m,
\label{eq:critical-conditions}
\end{equation}
and, for profiles with two distinct turning points,
\begin{equation}
\hat\beta_{\min}^{\mathrm{out}}=\hat\beta_m.
\label{eq:boundary-order}
\end{equation}
The first condition changes the winding sign adjacent to the
lower boundary, while the second describes the merger of
interior turning points. Equation~\eqref{eq:boundary-order}
changes the ordering of inverse-temperature levels without
merging the extrema.

\begin{figure*}[t]
\centering
\includegraphics[width=\textwidth]{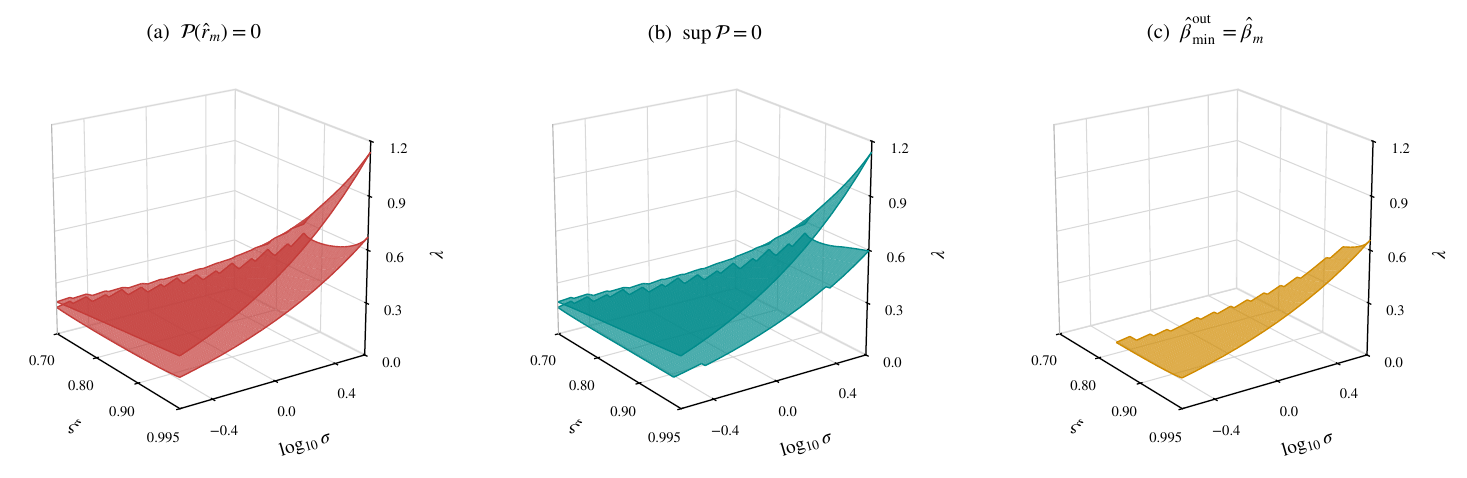}
\caption{Critical conditions in the explored parameter window.
Panels (a)--(c) show
$\mathcal P(\hat r_m)=0$,
$\sup_{\hat r_+\geq\hat r_m}\mathcal P(\hat r_+)=0$, and
$\hat\beta_{\min}^{\mathrm{out}}=\hat\beta_m$,
respectively. In panel (b), boundary maxima overlap with
panel (a), whereas interior maxima satisfy
$\mathcal P=\mathcal P'=0$.
The surfaces are obtained by numerical interpolation.}
\label{fig:parameter}
\end{figure*}

\begin{figure*}[t]
\centering
\includegraphics[width=\textwidth]{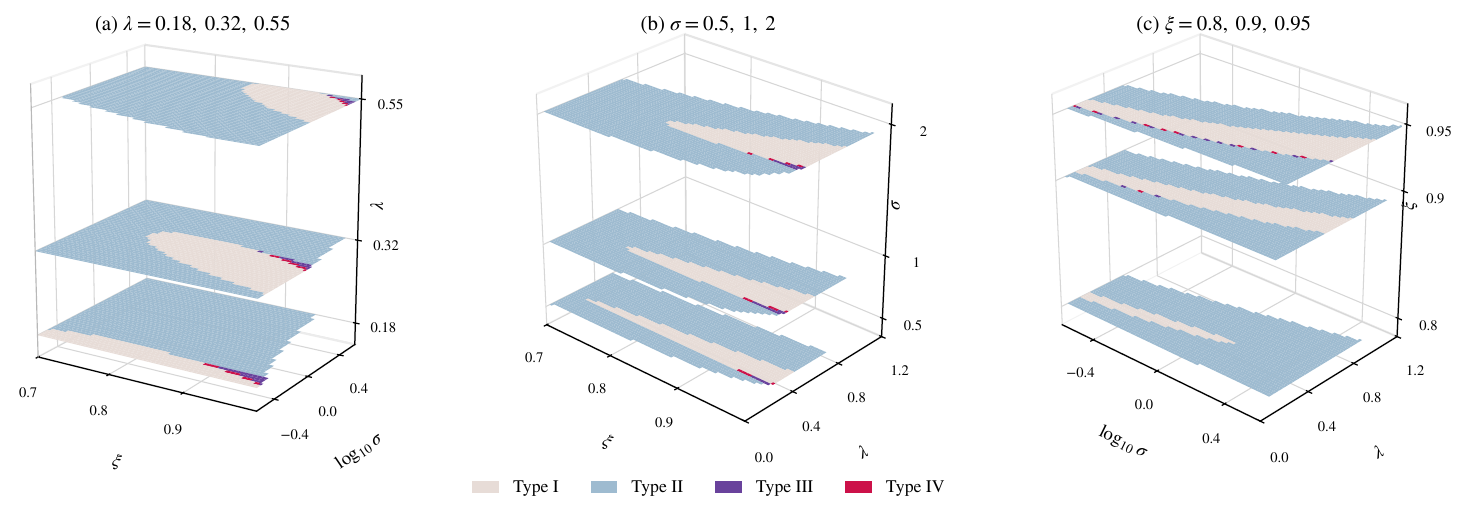}
\caption{Representative parameter sections at
(a) $\lambda=0.18,\,0.32,\,0.55$,
(b) $\sigma=0.5,\,1,\,2$, and
(c) $\xi=0.8,\,0.9,\,0.95$.
The four categories correspond to Types I--IV, as defined in the text.
Each regular grid point is classified from its computed turning points
and endpoint ordering.  Uncolored cells lie outside the displayed
endpoint window or correspond to critical configurations.}
\label{fig:sigma}
\end{figure*}

We identify representative regions by numerically solving the endpoint
condition~\eqref{eq:endpoint} and locating the interior zeros of
$\mathcal P$.  The parameter maps in Figs.~\ref{fig:parameter} and
\ref{fig:sigma} use the restricted window
$0.70\leq\xi\leq0.995$, $0.25\leq\sigma\leq4$, and
$0.025\leq\hat r_m\leq2.5$.  A $38\times15\times58$ grid samples
these variables, with logarithmic spacing in $\sigma$ and $\hat r_m$.
For each fixed $(\xi,\sigma,\hat r_m)$, the value of $\lambda$ is
determined by Eq.~\eqref{eq:endpoint}; it is not independently varied
along a defect curve.  The maps resolve four representative regions,
denoted Types I--IV.  Critical configurations are identified separately
from regions with simple turning points.
The resulting Types I--IV are therefore representative realizations within the explored parameter window rather than an exhaustive classification of the complete parameter space.

\subsection{Four representative branch structures}

The four representative parameter sets are given in
Table~\ref{tab:points}.  They are specified by the independent inputs
$(\xi,\sigma,\hat r_m)$, with $\lambda$ obtained from the physical
horizon endpoint condition.  The corresponding dimensionless defect
curves follow from Eq.~\eqref{eq:ell}, without further adjustment of
the radial domain.  Since $\sigma=1$ for these examples, the same
expression can be written as
\begin{align}
 \hat\beta_H^{(i)}(\hat r_+)&=
 \frac{\hat r_+}{1-\hat r_+^2
   [a_iU(\hat r_+)+c_iV(\hat r_+)]},\nonumber\\
 U(\hat r_+)&=(1+\hat r_+^{-4})^{2/3}-1,\nonumber\\
 V(\hat r_+)&=(1+\hat r_+^{-4})^{1/2}-1,\nonumber\\
 a_i&=\frac{3\lambda_i(1-\xi_i)}{\sqrt2},\qquad
 c_i=2\sqrt2\lambda_i\xi_i.
 \label{eq:profiles-normalized}
\end{align}
In each case the physical branch begins at
$\hat r_+=\hat r_m$, where $M_0=0$, and extends to infinity.
The endpoint solution is included in the geometric family but is not
counted as an interior thermodynamic defect.

\begin{table*}[t]
\centering
\caption{Parameters and turning-point values for the four representative
profiles.  The first three numerical columns are independent inputs;
$\lambda$ and all inverse temperatures are calculated from
Eqs.~\eqref{eq:endpoint} and \eqref{eq:ell}.
A dash indicates that the corresponding interior extremum is absent.
Numbers are rounded for display.}
\label{tab:points}
\begin{ruledtabular}
\begin{tabular}{lccccccc}
Type & $\xi$ & $\sigma$ & $\hat r_m$ & $\lambda$ &
$\hat\beta_m$ & $\hat\beta_{\min}^{\mathrm{out}}$ & $\hat\beta_{\max}$\\
\hline
I   & 0.9500 & 1 & 0.5000 & 0.335839 & 2.0370 & 1.6488$^{a}$ & ---\\
II  & 0.9000 & 1 & 0.1000 & 0.182139 & 0.2762 & --- & ---\\
III & 0.9751 & 1 & 0.1808 & 0.310242 & 1.4928 & 1.5434 & 1.6554\\
IV  & 0.9837 & 1 & 0.1496 & 0.319534 & 1.6322 & 1.5821 & 1.8340
\end{tabular}
\end{ruledtabular}
\par\smallskip
\footnotesize $^{a}$For Type I this value denotes its single minimum,
$\hat\beta_{\min}$, rather than an outer minimum of a two-turning-point
profile.
\end{table*}

\begin{figure*}[t]
\centering
\includegraphics[width=\textwidth]{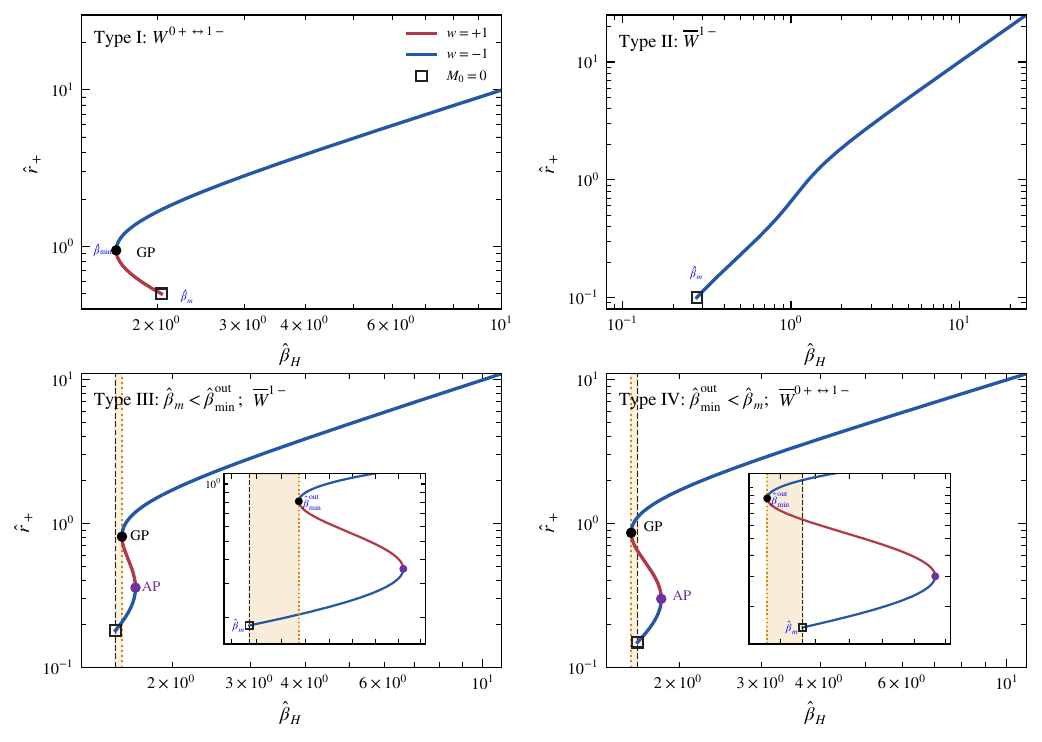}
\caption{Zero points of the vector field in the
$\hat\tau$--$\hat r_+$ plane for the four representative profiles:
Type I, Type II, Type III, and Type IV.
Red and blue curves represent branches with $w=+1$ and $w=-1$,
respectively.  Open squares locate the $M_0=0$ physical lower
endpoint.
Marked interior extrema are labeled as generation points (GP) or annihilation points (AP).
The enlarged views distinguish the orderings
$\hat\beta_m<\hat\beta_{\min}^{\mathrm{out}}$ (Type III) and
$\hat\beta_{\min}^{\mathrm{out}}<\hat\beta_m$ (Type IV).}
\label{fig:profiles}
\end{figure*}

All four profiles obey $0<\hat\beta_m<\infty$ and
$\hat\beta_H(\infty)=\infty$, as established in Sec.~IV.
They are distinguished by the sign of $\mathcal P$ adjacent to
$\hat r_m$, the number of simple interior turning points, and the
ordering of the associated inverse-temperature levels.
The resulting branch organization is shown in Fig.~\ref{fig:profiles}.

\paragraph*{\textbf{Type I: $W^{0+\leftrightarrow1-}$.}}
The defect curve initially decreases from $\hat\beta_m$ to a single
minimum $\hat\beta_{\min}<\hat\beta_m$ and subsequently increases
to infinity.  The radial winding sequence is $[+,-]$.
A pair with winding numbers $(+1,-1)$ is generated at
$\hat\tau=\hat\beta_{\min}$.  The positive-winding branch reaches
the lower boundary at $\hat\tau=\hat\beta_m$; beyond this level only
the outer negative-winding defect remains.  The transition changes
$W$ from $0$ to $-1$, as in the previously proposed
$W^{0+\leftrightarrow1-}$ structure~\cite{2025PhRvD.111f1501W}.

\paragraph*{\textbf{Type II: $\overline W^{1-}$.}}
The inverse-temperature curve is strictly increasing on the counted
radial domain.  Every finite thermodynamic defect carries $w=-1$.
No defect exists for $\hat\tau<\hat\beta_m$; one negative-winding
defect is present for $\hat\tau>\hat\beta_m$.
There are no interior generation or annihilation points.  This is
the monotonic realization of the proposed $\overline W^{1-}$
pattern~\cite{2025PhRvD.111f1501W}.

\paragraph*{\textbf{Type III: $\overline W^{1-}$.}}
The curve has a local maximum at $\hat r_a$ and a subsequent local
minimum at $\hat r_b$, with
\begin{equation}
 \hat\beta_m<\hat\beta_{\min}^{\mathrm{out}}
 <\hat\beta_{\max}.
 \label{eq:typeIII-order}
\end{equation}
The radial winding sequence is $[-,+,-]$.
A negative-winding defect first enters through the lower boundary
at $\hat\tau=\hat\beta_m$.  At
$\hat\tau=\hat\beta_{\min}^{\mathrm{out}}$, a pair of defects with
opposite winding numbers is generated on the intermediate and outer
segments.  At $\hat\tau=\hat\beta_{\max}$, the inner negative-winding
and intermediate positive-winding branches annihilate,
leaving the outer negative-winding branch.
Thus one generation and one annihilation point modify the number of
defects without changing $W=-1$ on any nonempty regular slice.
Type III provides a nonmonotonic realization of
$\overline W^{1-}$,
which was introduced as an extended subclass in
thermodynamic topology classifications
\cite{2025PhRvD.111f1501W}.

\paragraph*{\textbf{Type IV: $\overline W^{0+\leftrightarrow1-}$.}}
This profile also has the radial winding sequence $[-,+,-]$, but
its inverse-temperature levels satisfy the opposite boundary ordering,
\begin{equation}
 \hat\beta_{\min}^{\mathrm{out}}
 <\hat\beta_m<\hat\beta_{\max}.
 \label{eq:typeIV-order}
\end{equation}
At $\hat\tau=\hat\beta_{\min}^{\mathrm{out}}$, a positive-winding
intermediate defect and a negative-winding outer defect are generated.
The resulting winding sequence on
$\hat\beta_{\min}^{\mathrm{out}}<\hat\tau<\hat\beta_m$ is
$[+,-]$, with $W=0$.  As $\hat\tau$ passes through
$\hat\beta_m$, a negative-winding defect enters through the
physical lower boundary, producing the three-defect sequence
$[-,+,-]$ and changing $W$ to $-1$.  At
$\hat\tau=\hat\beta_{\max}$, the inner negative-winding and
intermediate positive-winding branches annihilate; the remaining
outer branch has $w=-1$.  Hence the nonempty winding sequences evolve
as
\begin{equation}
 \begin{aligned}
 (w_i):&\quad [+,-]\longrightarrow[-,+,-]\longrightarrow[-],\\
 W:&\quad 0\longrightarrow-1\longrightarrow-1.
 \end{aligned}
 \label{eq:typeIV-evolution}
\end{equation}
This evolution motivates the candidate thermodynamic topological
subclass $\overline W^{0+\leftrightarrow1-}$.
For the representative parameters in Table~\ref{tab:points},
$\mathcal P'(\hat r_a)\simeq0.3024$ and
$\mathcal P'(\hat r_b)\simeq-0.5834$.
By the implicit function theorem and continuity, the two simple
turning points and the strict ordering in
Eq.~\eqref{eq:typeIV-order} persist in an open parameter
neighborhood.

For Type III,
the two turning radii are
$\hat r_a=0.3584$ and $\hat r_b=0.8106$, with
$(\hat\beta_{\max},\hat\beta_{\min}^{\mathrm{out}})
=(1.6554,1.5434)$.  The corresponding Type IV values are
$\hat r_a=0.2997$ and $\hat r_b=0.8623$, with
$(\hat\beta_{\max},\hat\beta_{\min}^{\mathrm{out}})
=(1.8340,1.5821)$. Types III and IV have the same radial winding sequence and
two interior extrema. Their outer minimum and $\hat\beta_m$
occur in the opposite order.

Table~\ref{tab:counts} summarizes the defect evolution of the four representative profiles.
The radial winding
sequences are ordered by increasing horizon radius, while the nonempty
defect configurations are listed in order of increasing auxiliary
inverse temperature.

\begin{table*}[t]
\centering
\caption{Thermodynamic topological characteristics of the four
representative Einstein--NLED morphologies. The boundary-field directions $I_1$--$I_4$ refer to the outer
radial, upper angular, inner radial, and lower angular boundaries;
$I_3=\leftarrow/\rightarrow$ corresponds to
$\hat\tau<\hat\beta_m$/$\hat\tau>\hat\beta_m$.
Common inverse-temperature asymptotic behavior:
$0<\beta_H(r_m^{+})<\infty,\qquad\beta_H(\infty)=\infty$
}
\label{tab:counts}
\begingroup
\scriptsize
\setlength{\tabcolsep}{3pt}
\renewcommand{\arraystretch}{1.25}
\begin{ruledtabular}
\begin{tabular}{llcccccccc}
Type & Class/subclass & $(w_i)_{\rm radial}$ & $\Delta\hat\beta$
& Nonempty $(w_i)$ & $W$ & $I_1$ & $I_2$ & $I_3$ & $I_4$ \\
\hline
I & $W^{0+\leftrightarrow1-}$ & $[+,-]$ & ---
  & $[+,-]\to[-]$ & $0\to-1$
  & $\leftarrow$ & $\uparrow$ & $\leftarrow/\rightarrow$ & $\downarrow$ \\
II & $\overline W^{1-}$ & $[-]$ & ---
  & $[-]$ & $-1$
  & $\leftarrow$ & $\uparrow$ & $\leftarrow/\rightarrow$ & $\downarrow$ \\
III & $\overline W^{1-}$ & $[-,+,-]$ & $>0$
  & $[-]\to[-,+,-]\to[-]$ & $-1$
  & $\leftarrow$ & $\uparrow$ & $\leftarrow/\rightarrow$ & $\downarrow$ \\
IV & $\overline W^{0+\leftrightarrow1-}$ & $[-,+,-]$ & $<0$
  & $[+,-]\to[-,+,-]\to[-]$ & $0\to-1$
  & $\leftarrow$ & $\uparrow$ & $\leftarrow/\rightarrow$ & $\downarrow$ \\
\end{tabular}
\end{ruledtabular}
\endgroup
\end{table*}

\subsection{Local winding numbers and the boundary degree}

The winding numbers follow locally from the Jacobian in
Eq.~\eqref{eq:jac}.  They can also be obtained directly from the
mapping of a counterclockwise contour enclosing an isolated zero
$Z_i=(\hat r_i,\pi/2)$.  For a contour contained in the physical
thermodynamic parameter space, set
\begin{align}
 \hat r_+(\vartheta)&=\hat r_i+\epsilon_r\cos\vartheta,
 \qquad
 \Theta(\vartheta)=\frac{\pi}{2}+\epsilon_\Theta\sin\vartheta,
 \nonumber\\
 \Omega_i(\vartheta)&=\operatorname{unwrap}\arg
 \bigl(\phi^{r_+}+i\phi^\Theta\bigr),\nonumber\\
 w_i&=\frac{\Omega_i(2\pi)-\Omega_i(0)}{2\pi}.
 \label{eq:contour}
\end{align}
The contour is chosen to contain a single simple zero and not to
intersect the physical lower boundary.  Its image winds once around
the origin of the $\boldsymbol\phi$ plane, with orientation fixed by
the sign of $J_i$. 
The vector-field configurations and contour mappings are
illustrated for Types I and IV, representing the previously
proposed $W^{0+\leftrightarrow1-}$ pattern and the proposed
$\overline W^{0+\leftrightarrow1-}$ subclass, respectively.

\begin{figure*}[t]
\centering
\includegraphics[width=\textwidth]{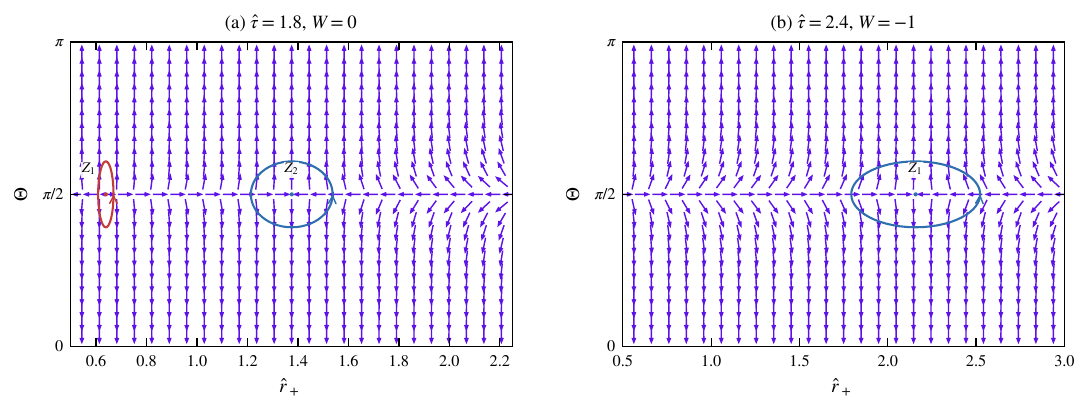}
\caption{Vector field configuration around thermodynamic defects
for Type I at $\hat\tau=1.8$ and $2.4$.  The zero points at
$\Theta=\pi/2$ carry the winding sequences $(+1,-1)$ and $(-1)$,
respectively.}
\label{fig:vectorA}
\end{figure*}

\begin{figure*}[t]
\centering
\includegraphics[width=\textwidth]{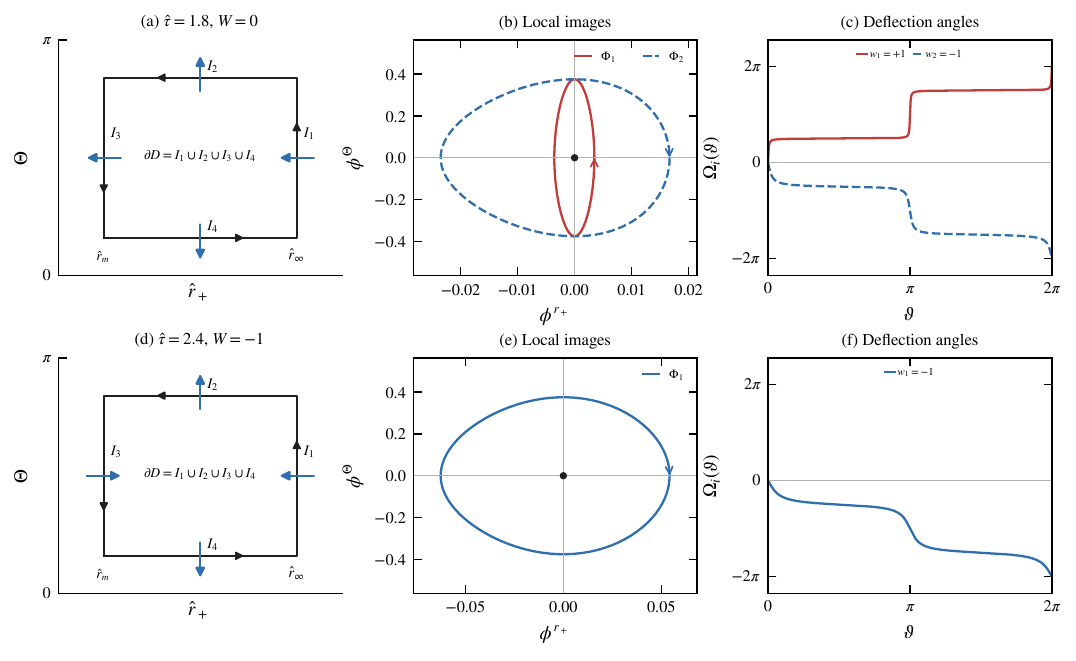}
\caption{Mapping of the vector field along closed contours enclosing
the Type I defects.  The contour images in $\boldsymbol\phi$ space
and accumulated angles $\Omega_i(\vartheta)$ verify the winding
numbers of Fig.~\ref{fig:vectorA}.}
\label{fig:windingA}
\end{figure*}

\begin{figure*}[t]
\centering
\includegraphics[width=\textwidth]{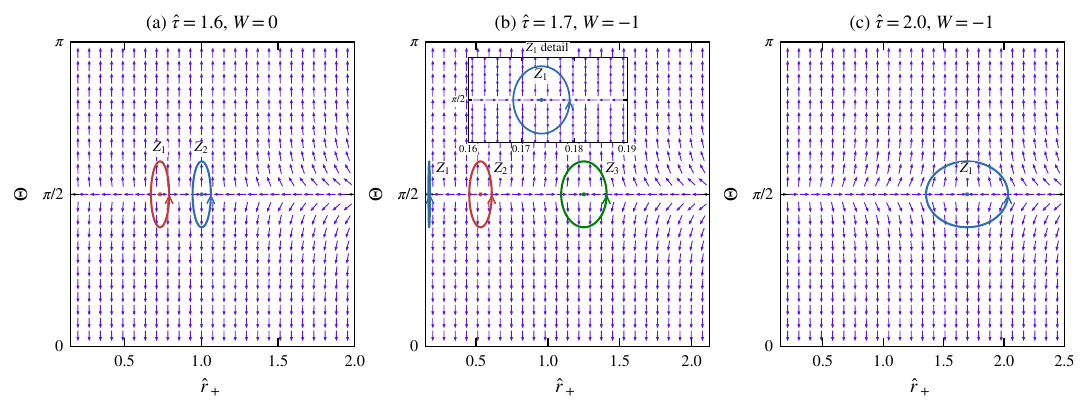}
\caption{Vector field configuration around thermodynamic defects
for Type IV at $\hat\tau=1.6$, $1.7$, and $2.0$.
The respective numbers of interior zeros are two, three, and one,
with $W=0,-1,-1$.}
\label{fig:vectorB}
\end{figure*}

\begin{figure*}[t]
\centering
\includegraphics[width=\textwidth]{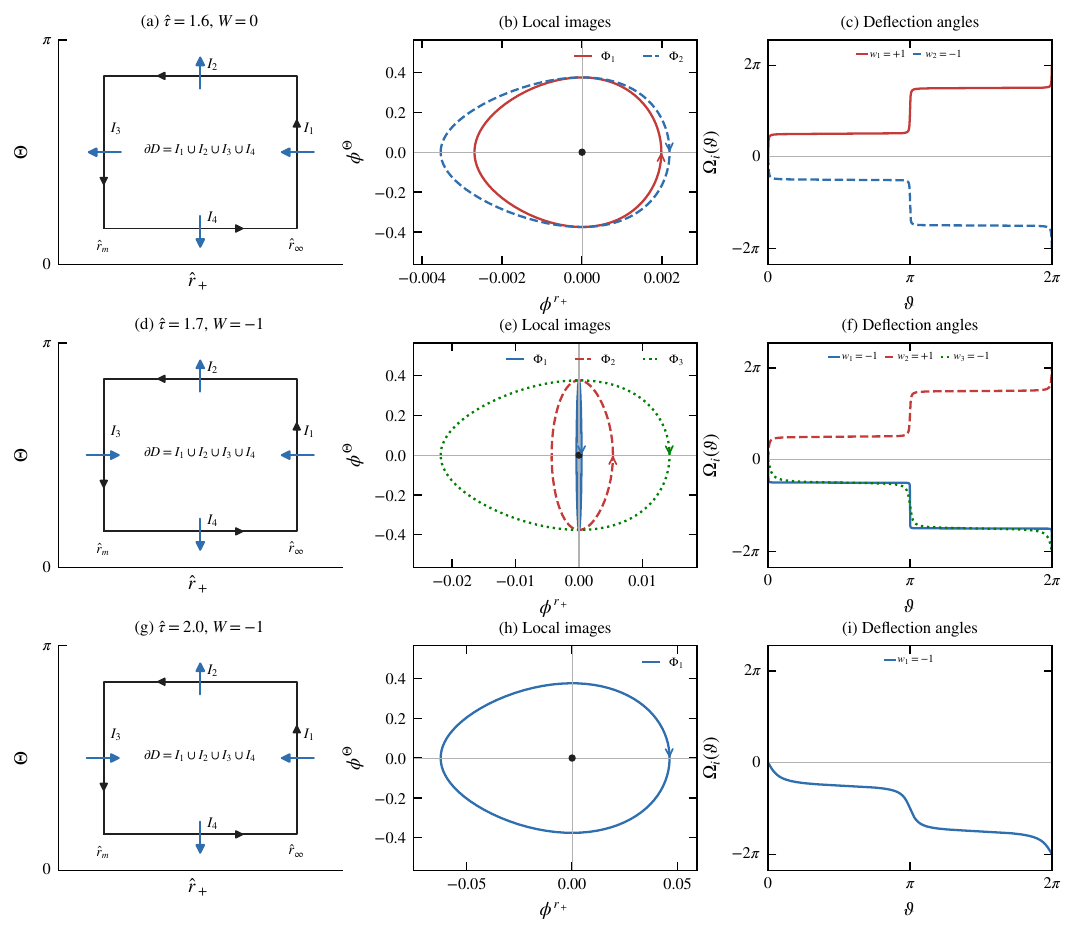}
\caption{Mapping of the vector field along closed contours enclosing
Type IV defects.  In the three-defect slice the ordered accumulated
angles change by $-2\pi$, $+2\pi$, and $-2\pi$, verifying the winding
sequence $[-,+,-]$.}
\label{fig:windingB}
\end{figure*}

For Type I, the representative slices $\hat\tau=1.8$ and $2.4$
contain, respectively, the winding sequences $(+1,-1)$ and $(-1)$.
The corresponding vector fields and contour images are displayed in
Figs.~\ref{fig:vectorA} and \ref{fig:windingA}.  For Type IV, the
slices $\hat\tau=1.6,1.7,2.0$ contain the sequences
$(+1,-1)$, $(-1,+1,-1)$, and $(-1)$, as shown in
Figs.~\ref{fig:vectorB} and \ref{fig:windingB}.  

The change in the global topological number can be established
without restricting attention to a particular representative profile.
At fixed regular $\hat\tau$, consider the vector field on the
rectangle
\begin{equation}
 D_{\epsilon,\delta,R}=
 (\hat r_m+\epsilon,R)\times(\delta,\pi-\delta),
 \label{eq:domain-degree}
\end{equation}
where $\epsilon,\delta>0$ are small and $R$ is sufficiently large
to enclose all finite zeros.  Provided the field is nonzero on the
boundary, Duan's $\phi$-mapping construction gives
\begin{equation}
 \deg(\boldsymbol\phi,D_{\epsilon,\delta,R},0)
 =\sum_i w_i.
 \label{eq:boundary-degree}
\end{equation}
The angular component has a single zero at $\Theta=\pi/2$ with
positive derivative.  Accordingly, the signed count of the full
vector field reduces to the signed crossings of
$g(\hat r_+)=\phi^{r_+}(\hat r_+,\hat\tau)$ along the radial
interval.  For simple zeros,
\begin{equation}
 W(\hat\tau)=\frac{1}{2}
 \left[\operatorname{sgn}g(\infty)
       -\operatorname{sgn}g(\hat r_m^+)\right].
 \label{eq:degree-crossings}
\end{equation}
The two endpoint signs are determined by Eq.~\eqref{eq:ell}:
\begin{align}
 g(\hat r_m^+)&=\frac{\hat r_m}{2}
 \left(\frac{1}{\hat\beta_m}-\frac{1}{\hat\tau}\right),
 \nonumber\\
 g(\hat r_+)&=-\frac{\hat r_+}{2\hat\tau}+\frac12
 +O(\hat r_+^{-2}),\qquad \hat r_+\to\infty.
 \label{eq:radial-boundary}
\end{align}
Consequently, $g(\infty)<0$, whereas the lower boundary has the
sign of $\hat\tau-\hat\beta_m$.  
The relation between bulk winding numbers and boundary contributions
has been discussed in the broader $\phi$-mapping framework and recent
thermodynamic topology studies
\cite{1998NuPhB.514..705D,2026arXiv260707364W}.
The global winding sum is therefore
\begin{equation}
 W(\hat\tau)=
 \begin{cases}
 0,  &0<\hat\tau<\hat\beta_m,\\
 -1, &\hat\tau>\hat\beta_m,
 \end{cases}
 \label{eq:degree}
\end{equation}
for regular slices not containing a degenerate interior zero.
At $\hat\tau=\hat\beta_m$, a zero reaches the physical lower
boundary of the counted domain, so the boundary-nonvanishing
condition required for the limiting degree fails.
Equation~\eqref{eq:degree} applies to all four representative
profiles. Interior generation and annihilation of
opposite-winding defect pairs leave $W$ unchanged
\cite{2026arXiv260707364W}.

\subsection{Numerical survey and scope of the classification}

To examine the parameter dependence of the four representative
branch structures, we performed a numerical survey in
$(\xi,\sigma,\lambda)$ over
$0.001\leq\xi\leq0.999$,
$10^{-3}\leq\sigma\leq10^3$, and
$0.005\leq\lambda\leq2$.
A $120\times120\times200$ tensor-product grid was used,
with linear sampling in $\xi$ and $\lambda$ and logarithmic
sampling in $\sigma$.
For each parameter set, the physical lower endpoint was
determined from $M_0(r_m)=0$, and the interior turning points
were identified from $\mathcal P(\hat r_+)=0$ on
$\hat r_+>\hat r_m$.
The survey recorded $28\,680$ Type I, $2\,849\,785$ Type II,
$1\,140$ Type III, and $395$ Type IV configurations.
No additional regular morphology was identified within
the sampled parameter region.

A useful analytic restriction follows from differentiating
$\mathcal P$:
\begin{align}
 \mathcal P'(\hat r_+)
 &=2\sqrt2\lambda\hat r_+
 [-\ell-2y\ell_y-8y^2\ell_{yy}],
 \qquad y=\hat r_+^{-4},\nonumber\\
 \mathcal P(\hat r_+)&\leq
 -1+\frac{4\sqrt2\lambda}{\hat r_+^2}.
 \label{eq:Pprime}
\end{align}
The inequality uses $0<\ell_y\leq1$ and $\ell\geq0$ for the
NLED Lagrangian under consideration.  In particular,
$\mathcal P\leq-1/2$ for
$\hat r_+\geq\sqrt{8\sqrt2\lambda}$,
which excludes additional turning points beyond this radius.
This bound controls the large-radius search region, while the
numerical results establish the four reported morphologies only
within the explored parameter domain.

\subsection{Comparison with existing thermodynamic topological (sub)classes}

\begin{table*}[t]
\centering
\caption{Selected published thermodynamic topological sub(classes) compared with the new (sub)classes identified in this work.}
\label{tab:topology-comparison}
\begingroup
\scriptsize
\renewcommand{\arraystretch}{1.20}
\begin{ruledtabular}
\begin{tabular*}{\textwidth}{@{\extracolsep{\fill}}lcccccc}
Classes & Innermost & Outermost &
Low $T$ ($\beta_H\to\infty$) & High $T$ ($\beta_H\to0$) & DP & $W$ \\
\hline
$W^{0-\leftrightarrow1+}$
 & Unstable & Stable & No & Stable large & One More AP & $0$ or $+1$ \\
$\overline W^{1+}$
 & Stable & Stable & No & Stable large & In pairs & $+1$ \\
$\widehat W^{1+}$
 & Stable & Stable & \shortstack{Unstable small +\\two stable small}
 & Stable large & One More GP & $+1$ \\
$\widetilde W^{1+}$
 & Unstable & Stable & Stable small
 & \shortstack{Unstable small + stable small\\+ stable large}
 & One More AP & $+1$ \\
$\ddot W^{1-}$
 & Unstable & Stable & Unstable small
 & \shortstack{Two unstable small\\+ stable large}
 & One More AP & $-1$ \\
 \hline
$W^{0+\leftrightarrow1-}$
 & Stable & Unstable & Unstable large & No & One More GP & $0$ or $-1$ \\
$\overline W^{1-}$
 & Unstable & Unstable & Unstable large & No & In pairs & $-1$ \\
$\overline W^{0+\leftrightarrow1-}$ 
 & Unstable & Unstable & Unstable large & No & In pairs & $0$ or $-1$ \\
\end{tabular*}
\end{ruledtabular}
\endgroup
\end{table*}

The thermodynamic properties of the extended classes and subclasses
introduced in Refs.~\cite{2025PhRvD.111f1501W,2025PhRvD.112l4024A,
2025EPJC...85.1386C}, together with realizations in charged,
accelerating, and higher-curvature black holes
\cite{2023PhRvD.108h4041W,2024JHEP...06..213W,
2023PhRvD.107f4023L}, provide a basis for comparing the four
Einstein--NLED morphologies (see Table~\ref{tab:topology-comparison} for a summary). 

\textit{Wu} et al. identified topological class $W^{0-\leftrightarrow1+}$ and topological subclass $\overline{W}^{1+}$, and also conjectured topological class $W^{0+\leftrightarrow1-}$ and topological subclass $\overline{W}^{1-}$ from the symmetry of the $(r_h,\beta_H)$ profiles \cite{2025PhRvD.111f1501W}. The latter two share a finite positive lower-endpoint inverse temperature and an unstable large-black-hole branch at low temperature. They differ in their radial branches: $W^{0+\leftrightarrow1-}$ has a stable innermost and an unstable outermost branch, with one more GP than AP and $W=0$ or $-1$, indicating that black holes in this class exhibit temperature-dependent thermodynamic topological phase transitions; $\overline{W}^{1-}$ has an unstable innermost branch and an unstable outermost branch, equal numbers of GP and AP (including zero), and $W=-1$.

The corresponding defect curves are shown in Figs.~\ref{fig:profiles}(a)--(c). All three profiles have a finite positive inverse temperature at the lower horizon boundary and an unstable large-black-hole branch as $\beta_H\to\infty$; no defect exists in the high-temperature limit $\beta_H\to0$. Figure~\ref{fig:profiles}(a) realizes the conjectured
$W^{0+\leftrightarrow1-}$ class. No black-hole defect
exists in the high-temperature limit $\beta_H\to0$,
whereas a single unstable large black hole remains as
$\beta_H\to\infty$. At finite inverse temperature, a GP
generates a stable small and an unstable large black hole
at $\hat\tau=\hat\beta_{\min}$. Both exist for
$\hat\beta_{\min}<\hat\tau<\hat\beta_m$, with $W=0$.
As $\hat\tau$ passes $\hat\beta_m$, the stable small-black-hole
branch exits through the physical lower horizon boundary.
Only the unstable large-black-hole branch remains, and
$W$ changes to $-1$. This boundary-induced change is
the thermodynamic topological phase transition associated
with $W^{0+\leftrightarrow1-}$. 

Figures~\ref{fig:profiles}(b) and \ref{fig:profiles}(c)
provide two realizations of the conjectured
$\overline{W}^{1-}$ subclass. In both cases, no black-hole
defect exists in the high-temperature limit
$\beta_H\to0$, whereas a single unstable large black hole
remains in the low-temperature limit $\beta_H\to\infty$.
Type II has one unstable branch throughout its radial
domain: a defect appears near the lower horizon boundary
for $\hat\tau>\hat\beta_m$ and moves to larger radius as
$\hat\tau$ increases. Type III has a richer finite-temperature
structure. For
$\hat\beta_m<\hat\tau<\hat\beta_{\min}^{\mathrm{out}}$,
only an unstable small black hole exists. A GP then
produces a stable intermediate and an unstable large
black hole, giving $[-,+,-]$. At the AP, the unstable
small and stable intermediate branches annihilate,
leaving the unstable large black hole. Type II thus has
zero GP--AP pairs and Type III has one, while $W=-1$
on every nonempty regular slice.

Figure~\ref{fig:profiles}(d) displays Type IV. Its inverse
temperature rises from the finite positive value
$\hat\beta_m$ to a local maximum, falls to an outer minimum
below $\hat\beta_m$, and then diverges at large radius.
Thus no black-hole state exists in the high-temperature
limit, whereas an unstable large black hole remains in the
low-temperature limit. At
$\hat\beta_{\min}^{\mathrm{out}}<\hat\tau<\hat\beta_m$,
a stable smaller black hole and an unstable larger black
hole coexist, with $W=0$. When an unstable small black
hole enters through the lower horizon boundary, the states
become ordered as unstable small, stable intermediate, and
unstable large black holes, and $W$ changes to $-1$.
The unstable small and stable intermediate states subsequently
annihilate, leaving the unstable large black hole. Hence
the innermost and outermost black-hole branches are unstable
and the GP and AP occur in a pair, as predicted for
$\overline{W}^{1-}$; nevertheless, this boundary crossing gives a thermodynamic topological phase transition from $W=0$ to $W=-1$, as in topological class $W^{0+\leftrightarrow1-}$. Therefore, following the convention of ordering black-hole states
from small to large radii and accounting for its distinct
thermodynamic evolution, we propose to denote this subclass by
$\overline{W}^{0+\leftrightarrow1-}$.


\section{Discussion and conclusions}
\label{sec:6}
By viewing black-hole states as topological defects in thermodynamic
parameter space, we investigate the static magnetic black hole of
Einstein gravity coupled to causal nonlinear
electrodynamics~\cite{2026arXiv260806639F}. Using the off-shell
free energy and Duan's $\phi$-mapping topological current theory
\cite{1998NuPhB.514..705D}, we survey its physical single-horizon
sector with $M_0\geq0$ in the canonical ensemble. For each fixed choice of $(Q_m,\beta,\sigma,\xi)$, the zeros of
the off-shell vector field trace the equilibrium defect curve
$\tau=\beta_H(r_+)$. Scanning the admissible parameter space
then reveals distinct defect-curve profiles and thermodynamic
branch structures.

We identify four distinct defect-curve profiles. Three provide
black-hole realizations of the previously conjectured
$W^{0+\leftrightarrow1-}$ and $\overline W^{1-}$
(sub)classes~\cite{2025PhRvD.111f1501W}. The fourth reveals
a new thermodynamic topological subclass, which we denote by
$\overline W^{0+\leftrightarrow1-}$.
The new subclass has an unstable innermost and outermost branch.
Its inverse temperature is finite at the lower horizon boundary
and diverges at large radius, leaving no black hole at high
temperature and one unstable large black hole at low temperature.
As $\hat\tau$ increases, the states evolve from a stable--unstable
pair to an unstable--stable--unstable triple, then to one unstable
large black hole. The entry of the unstable small branch through
the lower horizon boundary changes $W:0\to-1$, giving a
thermodynamic topological phase transition.

It would be useful to test the proposed
$\overline W^{0+\leftrightarrow1-}$ subclass in other admissible
nonlinear-electrodynamic or gravitational models and to examine
whether changing the thermodynamic ensemble modifies its
boundary-induced defect evolution. The symmetry of the
$(r_h,\beta_H)$ profiles considered in
Ref.~\cite{2025PhRvD.111f1501W} also suggests a possible mirrored
pattern, provisionally denoted
$\overline W^{0-\leftrightarrow1+}$. A physically admissible black-hole realization of this mirrored
pattern has yet to be demonstrated.

The causal nonlinear-electrodynamics black-hole family thus provides
a setting for studying how physical horizon boundaries and defect
branches shape thermodynamic topology, including boundary-induced
thermodynamic topological phase transitions. Whether other admissible parameter regimes of this family, or
different black-hole solutions, exhibit further thermodynamic
topological (sub)classes remains an open question.

\begin{acknowledgments}
This research was  supported by the National Natural Science Foundation of China (Grant No.12265007), and the Guizhou Provincial Major Scientific and Technological Program (XKBF(2025)010).
\end{acknowledgments}

\bibliography{references}
\end{document}